\documentclass[aps,pra,physrev,amsmath,amssymb,reprint,superscriptaddress]{revtex4-1}
\usepackage[utf8]{inputenc}
\usepackage{CJKutf8}

\usepackage[utf8]{inputenc}
\usepackage{graphicx}
\usepackage[caption=false]{subfig}
\usepackage{braket}
\usepackage{natbib}
\renewcommand{\citet}[1]{Ref.~\cite{#1}}
\usepackage{color}
\usepackage{xcolor}
\usepackage{quantikz} 
\usepackage{adjustbox} 
\usepackage[version=4]{mhchem} 

\begin{document}

\title{A variational quantum eigensolver for dynamic correlation functions}
\begin{CJK*}{UTF8}{gbsn}
\author{Hongxiang Chen (陈鸿翔)}
\email{hongxiang.chen@rahko.ai}
\affiliation{Department of Computer Science, University College London}
\affiliation{Rahko Ltd., Finsbury Park,  N4 3JP, United Kingdom}
\author{Max Nusspickel}
\affiliation{Department of Physics, King's College London, Strand, London, WC2R 2LS, U.K.}
\author{Jules Tilly}
\affiliation{Department of Physics, University College London}
\affiliation{Rahko Ltd., Finsbury Park,  N4 3JP, United Kingdom}
\author{George H. Booth}
\email{george.booth@kcl.ac.uk}
\affiliation{Department of Physics, King's College London, Strand, London, WC2R 2LS, U.K.}

\date{\today}

\begin{abstract}
Recent practical approaches for the use of current generation noisy quantum devices in the simulation of quantum many-body problems have been dominated by the use of a variational quantum eigensolver (VQE). These coupled quantum-classical algorithms leverage the ability to perform many repeated measurements to avoid the currently prohibitive gate depths often required for exact quantum algorithms, with the restriction of a parameterized circuit to describe the states of interest. In this work, we show how the calculation of zero-temperature dynamic correlation functions defining the linear response characteristics of quantum systems can also be recast into a modified VQE algorithm, which can be incorporated into the current variational quantum infrastructure. This allows for these important physical expectation values describing the dynamics of the system to be directly converged on the frequency axis, and they approach exactness over all frequencies as the flexibility of the parameterization increases. The frequency resolution hence does not explicitly scale with gate depth, which is approximately twice as deep as a ground state VQE. We apply the method to compute the single-particle Green's function of {\em ab initio} dihydrogen and lithium hydride molecules, and demonstrate the use of a practical active space embedding approach to extend to larger systems. While currently limited by the fidelity of two-qubit gates, whose number is increased compared to the ground state algorithm on current devices, we believe the approach shows potential for the extraction of frequency dynamics of correlated systems on near-term quantum processors.
\end{abstract}
\maketitle
\end{CJK*}

\section{Introduction}
The accurate and efficient simulation of quantum many-body systems is a key challenge in computational physics, at the heart of fields as diverse as quantum chemistry, materials science, and quantum information. Open problems in the simulation of correlated materials include frustrated magnets \cite{Coey2010} and high-temperature superconductors \cite{Hasan2010}, for which the simulation of the interacting problem remains intractable to date despite significant recent progress. Dynamic correlation functions are central quantities in the understanding of quantum systems. These objects characterize the dynamics of quasiparticles through the (correlated) system, and are intimately tied to its resulting optical, magnetic, and transport properties. They also allow access to many static observables, notably including the total energy of the system. The single-particle Green's function, describing the propagation of a single electron or hole, is a particularly important example. This quantity, in its real-frequency representation, is directly accessible via (inverse) photoemission experiments, and is therefore a highly sought-after quantity in numerical simulations.

It is however a challenging quantity to compute, formally requiring access to the exponentially large set of eigenstates of the electron-attached and electron-removed systems. The Green's function has also taken on a more formal role within the Dynamical Mean-Field Theory~(DMFT), which at its heart requires the single-particle Green's function of a correlated impurity model. DMFT has emerged in the last couple of decades as one of the most promising approaches to describe strong correlation effects within a post density-functional theory framework \cite{Imada1998, Bednorz1986, Droghetti2017,Kovaleva2008, Weber2014}, thereby increasing the demand for methods that can accurately simulate the Green's function of correlated models.

A new frontier in the simulation of quantum systems has arisen from the advent of usable digital quantum computers. It has been argued since the days of Feynman that quantum systems should be used for the simulation of quantum problems, and while there are obvious advantages in this approach, it also requires the reformulation of algorithms for implementation on quantum devices, and for extraction of these quantities with the restricted (unitary) operation set intrinsic to quantum computation \cite{Lloyd1073, doi:10.1080/00268976.2011.552441, Cao2019, Bauer2020, mcardleQuantumComputationalChemistry2018}. While several methods such as quantum phase estimation (QPE) \cite{Cleve1998, Kitaev1996, Abrams1999} could provide exponential speed-up for the direct computation of the spectra of quantum systems, in practice, they require fault-tolerant quantum computing and is out of reach of current quantum devices. The current era of devices have been termed the noisy intermediate-scale quantum~(NISQ) computation\cite{Preskill2018}, where the emphasis for practical algorithms is on shallow gate depth and avoidance of significant numbers of auxiliary qubits to reduce the impact of uncorrected quantum gate/readout errors, decoherence, and paucity of coherent qubit arrays.

With these limitations in mind, the most applicable methods to date for current and near-term devices have relied on a hybrid quantum-classical optimization of the problem, termed the variational quantum eigensolver (VQE) \cite{VQE}. The VQE has also been extended into several methods to compute low lying excited states of systems in a state-specific fashion \cite{McClean2017, Colless2018, Santagati2018, Higgott2019, Jones2019, McArdle2019, Ollitrault2020, Tilly2020, zhangAdaptiveVariationalQuantum2021}. In these, a variational functional is found, which is sampled according to a parameterized quantum circuit, defining a wave function ansatz. The parameters of this circuit are then optimized classically in order to minimize the variational functional and find an approximation to the desired state. In general, this variational functional is the Ritz functional for the energy, and therefore minimizing the energy converges the circuit to an approximation to the ground state for a given Hamiltonian. With this method, the circuit can be parameterized to produce highly entangled quantum states (albeit with a compact number of parameters) which could not be efficiently sampled via classical approaches, such as the unitary coupled-cluster ansatz. It therefore enables accurate and efficient representations of many-body quantum states on quantum devices.

In this paper, we outline a modification of the VQE, which relies on a different variational functional that can be efficiently sampled, such that a parameterized state can be used on a quantum device to find arbitrary dynamical correlation functions of a given Hamiltonian. This is formulated directly in the frequency domain (either real or Matsubara frequency) and requires the optimization of a circuit defining the system correction vector at each desired frequency (although several optimization techniques help us avoiding to do a full computation at each frequency). This correction vector has previously been used for the optimization of frequency-domain equilibrium correlation functions for other parameterized states in electronic structure theory for `classical' computers, from neural quantum states \cite{ML_Corrvec} to tensor networks \cite{DMRG_Corrvec,DDMRG}, as well as being at the core of other electronic structure algorithms \cite{doi:10.1021/acs.jctc.8b00454, PhysRevB.101.045126,PhysRevB.85.205119}. It defines the response of the ground state to a perturbation at a given frequency, is prepared by an ansatz quantum circuit and optimized classically based on a sampled, variational cost function. The approach limits the requirement to a single ancillary qubit, and allows for relatively short circuit depths for use on near-term quantum devices.

The remainder of the paper is outlined as follows. In Sec.~\ref{sec:lit_rev} we present a review of relevant approaches for the extraction of equilibrium dynamical correlation functions on quantum devices. This is followed by a mathematical description of the proposed VQE for dynamic correlation functions (specifically the single-particle Green's function), with details of the required circuits for application on quantum devices. Furthermore, in Sec.~\ref{sec:qchem-cas} we discuss an approach to restrict the simulation of a correlated Green's function to an complete active orbital subspace, embedded in the mean-field response of low-energy unentangled orbitals. We present an approach to combining the subspace Green's functions rather than self-energies, in order to avoid the inversion of the noisy sampled active space Greens function matrix. In Sec.~\ref{sec:numerical_results} we numerically demonstrate the scheme on a quantum circuit simulator, applied to dihydrogen and lithium hydride with fully {\em ab~initio} correlated dynamics with interactions and quartic terms in the Hamiltonian. We introduce a noise model to mimic the two-qubit gate errors on near-term devices, which constitute the current bottleneck of the approach. Finally, Sec.~\ref{sec:discussion} provides a discussion on the implementation, advantages, and drawbacks of the method.

\section{Dynamical correlation functions on quantum devices} \label{sec:lit_rev}

There are a number of proposed approaches to access equilibrium dynamical correlation functions (and more specifically the equilibrium Green's function) on quantum devices, suitable both for longer-term and non-fault-tolerant near-term devices. These can be broadly categorized into whether they simulate the dynamics in the time or frequency domain.
For computing Green's function in the time domain, a number of approaches have been proposed based on Hamiltonian simulation following a Trotter-Suzuki \cite{Wecker2015, Bauer2016, Kreula2016, Kreula2016_,Keen2020}, or qubitization decomposition\cite{Low2019hamiltonian}. Another method based on a quantum generative model allows the computation of infinite temperature correlation functions \cite{PhysRevB.103.014301}.
While some of these methods have been successfully applied to impurity problems~\cite{Keen2020}, they generally require the gate depth (simulated time) to increase for a sufficiently fine energy resolution, and are therefore often out of reach of current devices.

A more noise-resilient method to compute Green's function in the time domain has been proposed in \citet{Endo2019}, which uses variational quantum simulation algorithm for the state evolution. Alternatively, imaginary time simulation of response properties has recently been demonstrated on current quantum devices \cite{PRXQuantum.2.010317}.

In the frequency domain (in which the current work is formulated), the computation of the Green's function is commonly recast as the solution to a quantum linear system problem \cite{PhysRevLett.103.150502}. These approaches generally rely on block encoding of the operator in order to represent it as a unitary, as well as amplitude amplification and phase estimation steps, and require a number of ancilla qubits and significant gate depths. They are therefore out of near-term capabilities of quantum computers. It is nonetheless worth noting that the convergence properties of these approaches depend on the condition number of the resulting operator and approaches to minimize the gate depth and apply the approach to dynamical quantities in quantum system have seen progress in recent years \cite{10.1145/3313276.3316366,doi:10.1137/16M1087072,PhysRevLett.122.060504,LinLin,Roggero2019, Kosugi2020, caiQuantumComputationMolecular2020}. Alternatively, QPE steps can be exchanged for a more NISQ-friendly VQE algorithm to calculate individual states or construct response functions \cite{DMFTIvan, Endo2019, zhuCalculatingGreenFunction2021, caiQuantumComputationMolecular2020}.

Here, we took a different approach which computes Green's function by solving the quantum linear system problem, but eliminates the requirement for the recasting of the operator into a unitary form. The approach extends the use of the variational quantum eigensolver (VQE) which leverages the quantum device in order to repeatedly sample an objective function of a parameterized correlated quantum state, and uses classical optimization strategies to iteratively refine that state. It employs a solution to the quantum linear system which requires only a single auxiliary qubit, has a gate depth which is only twice that of ground-state VQE algorithms, and is potentially more robust to noise on NISQ-era quantum devices compared with non-variational algorithms\cite{vqela}. A similar scheme has also recently been proposed in Ref.~\onlinecite{caiQuantumComputationMolecular2020}. Furthermore, the approach does not access individual states, and so exact spectral properties at a given frequency are obtained with increasing expressibility of the quantum ansatz.

\subsection{Green's functions}
\label{sec:method}

We first review the formulation of Green's functions in zero-temperature, time-independent fermionic systems, in order to formalize notation and terminology, and before describing the approach on quantum devices. We write the Hamiltonian in the grand-canonical ensemble as
\begin{equation} \label{eq:second_quant}
\mathcal{H}=\sum_{\alpha\beta} h_{\alpha\beta} c^\dagger_\alpha c_\beta 
+ \frac{1}{2} \sum_{\alpha\beta\gamma\delta} g_{\alpha\beta\gamma\delta} c^\dagger_\alpha c^\dagger_\beta c_\delta c_\gamma - \mu \sum_\alpha c^\dagger_\alpha c_\alpha,
\end{equation}
where $c_\alpha^{(\dagger)}$ are fermionic single-particle annihilation (creation) operators, $h_{\alpha\beta}$ and $g_{\alpha\beta\gamma\delta}$ are one- and two-body matrix elements between the degrees of freedom, $\mu$ is the chemical potential, and where $\alpha, \beta, \gamma, \delta$ are indices for the fermionic modes. The retarded Green's function~$G^\mathrm{R}(t)$ represents the linear response of the system to the addition and removal of an electron, defined as
\begin{equation} \label{eq:green_function}
G^\mathrm{R}_{\alpha\beta}(t) = -i \Theta(t)\langle \psi_0 | c_{\alpha}(t) c_{\beta}^{\dagger}(0)+c_{\beta}^{\dagger}(0) c_{\alpha}(t)| \psi_0 \rangle,
\end{equation}
where $c_{\alpha}(t)=e^{i\mathcal{H}t} c_{\alpha} e^{-i\mathcal{H}t}$ are fermionic operators in the Heisenberg representation, $\Theta(t)$ is the Heaviside step function, and $|\psi_0\rangle$ is the ground state for the Hamiltonian~$\mathcal{H}$ at a given chemical potential.

In order to formally evaluate the Fourier transform of Eq.~\ref{eq:green_function}, a regularization factor of $\lim_{\eta\to 0^+} e^{-\eta |t|}$ is included. Since $\eta$ is infinitesimal, this doesn't affect the Green's function at any physical, finite value of $t$. The regularized Fourier transform then takes the form
\begin{align} \label{eq:fourrier_trans}
G^\mathrm{R}_{\alpha\beta}(\omega) = \int_{-\infty}^{\infty} G^\mathrm{R}_{\alpha\beta}(t) e^{i\omega t - \eta |t|} \mathrm{d} t,
\end{align}
where the regularization can be understood as a small imaginary shift of the target frequency $\omega$. The real-frequency retarded Green's function is then obtained as
\begin{equation}
\begin{split}
G^\mathrm{R}_{\alpha\beta}(\omega) =& 
\langle \psi _{0} | c_{\alpha} \frac{1}{
\omega +i\eta + E_{0} -\mathcal{H}} c^{\dagger}_{\beta}|\psi _{0} \rangle \\
&+ \langle \psi _{0} |c^{\dagger}_{\beta} \frac{1}{
\omega +i\eta + \mathcal{H} -E_{0}} c_{\alpha}|\psi _{0} \rangle,
\end{split}
\end{equation}
where $E_0$ is the ground state energy of $\mathcal{H}$. It is often convenient to also consider the Green's function in imaginary time $\tau=it$,
\begin{align}
G_{\alpha\beta}( \tau ) =
&-\Theta ( \tau ) \langle \psi _{0} |c_{\alpha}( \tau ) c^{\dagger }_{\beta}| \psi _{0} \rangle 
\nonumber\\
&+\Theta ( -\tau ) \langle \psi _{0} |c^{\dagger }_{\beta} c_{\alpha}( \tau )| \psi _{0} \rangle ,
\end{align}
where $c_{\alpha}( \tau ) =e^{\tau \mathcal{H}} c_{\alpha} e^{-\tau \mathcal{H}}$. The Fourier transform to the imaginary-frequency domain defines the Matsubara Green's function, as
\begin{equation}
\begin{split}
G_{\alpha\beta}(i\omega_n) =&
\langle \psi _{0} |  c_{\alpha} \frac{1}{i\omega_n + E_{0} -\mathcal{H} } c^{\dagger}_{\beta} |\psi _{0} \rangle \\
&+\langle \psi _{0} |c^{\dagger}_{\beta} \frac{1}{i\omega_n + \mathcal{H} -E_{0} }c_{\alpha} |\psi _{0} \rangle ,
\end{split}
\end{equation}
where $\omega_n = (2n+1)\pi /\beta$ are the Matsubara frequencies,  $\beta$ is the inverse temperature, and $n\in\mathbb{Z}$. Note that since $G_{\alpha\beta} (i\omega_n) = G^{*}_{\alpha\beta} (-i\omega_n)$, $n$ can be restricted to $n\geqslant 0$.
It should be noted that the Matsubara frequencies form a continuous domain in the zero-temperature limit.

Both the real- and imaginary-frequency Green's function can be written in one general definition as
\begin{equation}
\label{eq:master_GF}
\begin{split}
G_{\alpha\beta} (z) =& 
\langle \psi _{0} | c_{\alpha} \frac{1}{z + E_{0} -\mathcal{H}} c^{\dagger}_{\beta} |\psi _{0} \rangle\\
&+\langle \psi _{0} | c^{\dagger}_{\beta} \frac{1}{z + \mathcal{H} - E_{0}} c_{\alpha} |\psi _{0} \rangle,
\end{split}
\end{equation}
such that $G^R(\omega)=G(\omega + i\eta)$ and the Matsubara Green's function is obtained by setting $z=i\omega_n$. From this form, we can define an operator $Q_{\pm}(z) = z \pm (\mathcal{H} -E_{0})$, with the sign depending on which term of Eq.~\ref{eq:master_GF} is being considered. 
By introducing $V_i$ as an operator defining the perturbation of the system for a general dynamic correlation function (and taking the form of $c_{\alpha/\beta}^{(\dagger)}$ for the single-particle Green's function), we can write the separate terms which contribute to the Green's function in the form
\begin{equation}
\begin{aligned}
\label{eq:G_ij}
G_{i j}(z) =  \langle \psi_0 | V_i^{\dagger} \frac{1}{Q(z)} V_j | \psi_0 \rangle 
=  \langle \psi_0 | V_i^{\dagger} | \chi_j(z) \rangle.
\end{aligned}
\end{equation}
The state $|\chi_j(z) \rangle$ defines the `correction vector' of the system, as
\begin{align}
\label{eq:def_chi_linear_eq}
|\chi_j(z) \rangle = \frac{1}{Q(z)} V_j| \psi_0 \rangle ,
\end{align}
and describes the linear response of the ground state to the dynamic perturbation at the chosen frequency. This correction vector is also found by many other algorithms in quantum simulation, including approaches in tensor networks and neural quantum states \cite{DMRG_Corrvec,DDMRG,ML_Corrvec}. We can further simplify the notation, by denoting the state $V_i|\psi_0\rangle $ as $|V_i\rangle$, such that once the correlation vector is obtained, the computation of $G_{ij}$ simplifies to $G_{ij}(z)=\langle V_i | \chi_j (z)\rangle$.

\subsection{Variational eigensolvers for correction vectors} \label{sec:corrvec}

In order to calculate $G_{ij}$ of Eq.~\ref{eq:G_ij}, we first map all fermionic operators to spin operators that can be directly measured on a quantum computer. Any operator can be mapped to a weighted sum of Pauli operators acting over the qubits, $\sum_{a} w_{a} P_{a}$, with $P_{a} \in \{X, Y, Z, I\}^{\otimes m}$, where $m$ is the number of qubits. There exist a number of possible mappings \cite{bravyi_fermionic_2002, Seeley2012, bravyi_tapering_2017, Setia2019, mcardleQuantumComputationalChemistry2018}; in this work we use the Jordan--Wigner mapping \cite{Jordan1928}. After this mapping, we use the VQE to find the ground state energy $E_0$ of a state represented by the parameterized quantum circuit, $U_{\mathrm{GS}}$\cite{VQE}. This circuit will prepare the ground state wave function $|\psi_0 \rangle$ from an initial state $\ket{0}$ of the quantum device as $\ket{\psi_0} = U_{\mathrm{GS}}\ket{0}$, though alternative ground state preparation algorithms could also be used \cite{QITE,Wang2019,yeteraydeniz2021benchmarking}.

We then aim to prepare a quantum circuit in order to solve for the correction vector via the linear equations $Q(z) |\chi_j(z)\rangle = |V_j\rangle$. In general, $|\chi_j(z)\rangle$ is not a normalized vector and therefore we rewrite it as $|\chi_j(z)\rangle =\gamma_j(z) |\tilde{\chi}_j(z)\rangle$, where $|\tilde{\chi}_j(z)\rangle$ is a normalized state and $\gamma_j(z) \in \mathbb{C}$ controls the true normalization of the correction vector for each frequency point ($z$) and perturbation (indexed by $j$). Our task is to find a second parameterized quantum circuit at each desired frequency point, $U_j(\theta)$, where $\theta$ represents the set of adjustable rotation angles in the chosen gates of the ansatz, such that the resulting circuit transforms the initial state $|0\rangle$ of the quantum device to $|\tilde{\chi}_j(z)\rangle$, i.e. $|\tilde{\chi}_j(z)\rangle=U_j(\theta(z))|0\rangle$. Once this is obtained, $\gamma_j(z)$ can be calculated as
\begin{align}
\label{eq:gamma}
\gamma_j(z) = \frac{\left\langle V_j|V_j \right\rangle}{\langle V_j | Q(z) U_j(\theta(z)) |0\rangle },
\end{align}
since at convergence
\begin{equation}
    |V_j\rangle = Q(z) |\chi_j \rangle = Q(z)\gamma_j(z) U_j(\theta)|0\rangle.
\end{equation}

To find $U_j(\theta)$, we follow a method similar to the variational linear equation resolution method presented in \citet{vqela}. We define a cost function, $g(\theta;z, j)$, as
\begin{align}
g(\theta;z, j) =& 
\langle 0 | U^\dagger_j(\theta) Q^\dagger(z) Q(z) U_j(\theta) |0\rangle  
\nonumber\\
& - \langle 0 | U^\dagger_j(\theta) Q^\dagger(z) \frac{ |V_j\rangle \langle V_j |}{\left\langle V_j|V_j \right\rangle} Q(z) U_j(\theta) |0\rangle \nonumber \\
=& 
\langle 0 | U^\dagger_j(\theta) Q^\dagger(z) Q(z) U_j(\theta) |0\rangle  
\nonumber\\
& - \frac{1}{\left\langle V_j|V_j \right\rangle} |\langle V_j | Q(z) U_j(\theta) |0\rangle|^2 \nonumber \\
=& \langle \psi | H'(z,j) |\psi \rangle, \label{eq:g}
\end{align}

where $|\psi \rangle = U_j(\theta) |0\rangle $ and $H'(z,j)$ is a new hermitian operator, defined as
\begin{equation}
\begin{aligned}
    H'(z,j) = Q^\dagger(z) \left(1 - \frac{|V_j\rangle \langle V_j |}{\left\langle V_j|V_j \right\rangle} \right) Q(z).
\end{aligned}
\end{equation}

This formalism is convenient, as $|\tilde{\chi}_j(z)\rangle$ is the normalized ground state of $H'(z,j)$, and thus $|\psi\rangle$ can be used as a variational approximation to it. When the cost function is minimized, the positive semi-definite projector $\left(1-\frac{|V_j\rangle \langle V_j |}{\left\langle V_j|V_j \right\rangle}\right)$ projects to the zero vector, as
\begin{align}
&\left(1 - \frac{|V_j\rangle \langle V_j |}{\left\langle V_j|V_j\right\rangle}\right) Q(z) |\tilde{\chi}_j(z)\rangle \nonumber \\
=&\,
\frac{1}{\gamma_j(z)}\left(1 - \frac{|V_j\rangle \langle V_j |}{\left\langle V_j|V_j\right\rangle}\right)  |V_j\rangle \nonumber \\
=&\, \vec{0},
\end{align}
ensuring that the minimal value of $g(\theta;z,j)=0$, corresponding to the lowest eigenvalue of $H'(z,j)$. Therefore, given a sufficiently flexible circuit parametrization, $U_j(\theta)$, minimizing $g(\theta;z,j)$ will yield the solution $\theta^*(z) =\mathrm{argmin}_{\theta} g(\theta;z,j)$, such that  $Q(z) U_j(\theta^*(z))|0\rangle \propto |V_j\rangle$, up to the constant factor $\gamma_j(z)$.
Hence, the final correction vector is found as
\begin{equation}
    |\chi_j\rangle=\gamma_j(z) Q(z) U_j(\theta^*(z))|0\rangle,
\end{equation}
with the value of $\gamma_j(z)$ at each frequency found via Eq.~\ref{eq:gamma}. 

\subsection{Implementation on a quantum device} \label{sec:implementation}

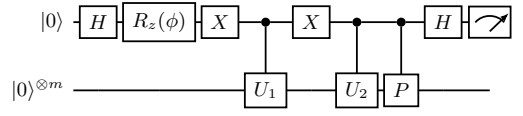
\begin{figure}
    \begin{center}
    \begin{adjustbox}{width=0.8\linewidth}
        \begin{quantikz}[column sep=1mm]
            & \lstick{$|0\rangle $} & \gate{H} & \gate{R_z(\phi)} & \gate{X}   & \ctrl{1}   & \gate{X} & \ctrl{1}   & \ctrl{1}& \gate{H} & \meter{} \\
            & \lstick{$|0\rangle^{\otimes m} $} & \qw      &\qw              & \qw        & \gate{U_1} & \qw      & \gate{U_2} & \gate{P} & \qw    & \qw
        \end{quantikz}
    \end{adjustbox}
    \end{center}
\caption{The quantum circuit used to compute overlaps of the form given in Eq.~\ref{eq:def_overlapterm}.
The upper horizontal line represents the ancillary qubit, with the lower line representing the $m$ qubits of the physical degrees of freedom.
The expectation value of the $Z$-measurement on the ancillary qubit yields 
$\mathrm{Re}\left( e^{i\phi}\langle 0 |U_1^\dagger P U_2 |0\rangle  \right)$ \cite{Endo2019}.
Hence, by choosing $\phi=0, \pi/2$, both the real and imaginary part of Eq.~\ref{eq:def_overlapterm} can be sampled.} 
\label{fig:overlap circuit}
\end{figure}

For a practical quantum implementation, we require an efficient way to sample from $g(\theta;z,j)$ at each desired $z$ and $V_j$, compute the norm of the correction vector $\gamma_j(z)$ from Eq.~\ref{eq:gamma}, and ultimately the overlap $\langle V_i|\chi_j(z)\rangle$ in order to compute $G_{ij}(z)$. 

Recall that both $V_{i}$ and $Q(z)$ have been mapped to a weighted sum of Pauli operators. 
Symmetric expectation values can therefore be computed in a straightforward fashion. These include e.g. the first term of Eq.~\ref{eq:g}, as
\begin{equation}\label{eq:expect_pauli_ops}
\begin{split}
&\langle 0|U^\dagger(\theta) Q^\dagger(z) Q(z) U(\theta)|0\rangle \\
=& \sum_{\alpha,\beta} w^*_{Q,a}w_{Q,b} \langle 0|U^\dagger(\theta) P_{a} P_{b} U(\theta)|0\rangle,
\end{split}
\end{equation}
since we can decompose $Q(z)$ into Pauli operator strings as $Q(z)=\sum_{a} w_{Q,a} P_{a}$ for some $w_{Q,a}\in \mathbb{C}$. Note that $P_{a} P_{b}$ is equal to another Pauli operator times a constant. Quantities in Eq.~\ref{eq:expect_pauli_ops} can be sampled in a straightforward manner in keeping with traditional VQE implementations \citep{VQE}. However, calculations of overlaps between different states, including the calculation of $\gamma_j(z)$, $G_{ij}(z)$, and the second contributing term to $g(\theta;z,j)$ in Eq.~\ref{eq:g}, require a more involved computation, similar to a Hadamard test circuit. All of these contributions can be decomposed into the form
\begin{equation}
\label{eq:def_overlapterm}
\left\langle 0 |U_1(\theta_1) P U_2(\theta_2) |0\right\rangle,
\end{equation}
where $U_1$ and $U_2$ are two (potentially parametrized) quantum circuits. We use the method described in \citet{Endo2019} to calculate the real and imaginary parts of Eq.~\ref{eq:def_overlapterm}, requiring the use of an additional single ancillary qubit, with the circuit shown schematically in Fig.~\ref{fig:overlap circuit}.
This circuit evolves the state as follows.
\begin{enumerate}
    \item After $H$ and $R_z(\phi)$ gates: 
    \begin{equation*}
    2^{-1/2}(|0\rangle +e^{i\phi}|1\rangle ) |0\rangle
    \end{equation*}
    \item After $X$, the controlled $U_1$, and $X$ gates:
    \begin{equation*}
        2^{-1/2}(|0\rangle U_1 |0\rangle +e^{i\phi}|1\rangle |0\rangle )
    \end{equation*}
    \item After the controlled $U_2$ and $P$ gates:
    \begin{equation*}
        2^{-1/2}(|0\rangle U_1 |0\rangle +e^{i\phi}|1\rangle P U_2 |0\rangle )
    \end{equation*}
    \item Before measurement: 
    \begin{equation*}\begin{split}
        2^{-1} |0\rangle \left( U_1|0\rangle  + e^{i\phi} P U_2 |0\rangle \right) &\\ + 2^{-1} |1\rangle  \left( U_1|0\rangle  - e^{i\phi} P U_2 |0\rangle  \right)
    \end{split}\end{equation*}
\end{enumerate}

Measurement of the ancillary qubit in the $Z$ basis can return $0$ or $1$, with probabilities $p(0)$ and $p(1)$ respectively. If $\phi=0$, then it can be verified that $p(0)-p(1)=\mathrm{Re} \langle 0| U_1^\dagger P U_2|0 \rangle$, while if $\phi=\pi$ it follows that $p(0)-p(1)=-\mathrm{Im}\langle 0|U_1^\dagger P U_2|0\rangle$. This approach allows us to compute all remaining quantities of interest and avoids the requirement of inversion of the $U_1$ circuit of the traditional Hadamard test.

\subsection{Active space embedding of Green's functions}
\label{sec:qchem-cas}

In order to overcome a restriction to small system sizes due to the limitations of qubit number, we can also seek a multiscale approach, whereby the quantum solution to the Green's function is solved in a strongly correlated subset of the total degrees of freedom, in the presence of a mean-field description of the response outside this. These multi-resolution or embedded approaches to the correlated dynamics are common in many approaches where the full solution is intractable, and where a subspace of the most important correlated physics can be identified, often involving subspaces chosen via locality or energetic arguments \cite{tilly2021reduced,Bauer2016,GalliQuantumEmbedding,yamazaki2018practical,dhawan2021dynamical,Takeshita2020,Yao2020,rossmannek2020quantum, Yalouz_2021}. In this work, we will demonstrate a simple approach to embed a noisy Green's function, e.g. sampled from a quantum device, within a mean-field environment described efficiently on a `classical' computer. 
This will employ a `complete active space' partitioning of the molecular orbitals of the {\em ab initio} system.

%
%
%
In this approach the ground state wave function and response vectors are expressed as the tensor product of a highly entangled quantum state within a complete active space~(CAS) of molecular orbitals, and a simple product state in the remaining `core' degrees of freedom, as
\begin{equation}
    \ket{\psi_0} = \ket{\psi_\mathrm{CAS}} \otimes \ket{\phi_\mathrm{core}} .
\end{equation}
The core state is therefore restricted to a single Slater determinant. In this work, the CAS space is simply chosen to span the highest-energy occupied and lowest-energy unoccupied single-particle orbitals which arise from a prior mean-field calculation. This space is therefore expected to span the most important low-energy quantum fluctuations required to describe the correlated physics of the system.
As a result of this approximation, many-body correlated states only have to be calculated in the CAS (in the presence of the static mean-field Coulomb and exchange potentials from the core electrons), reducing the number of qubits which are needed to describe a system.
%
%
Within the active space, an effective Hamiltonian is constructed according to
\begin{multline}\label{eq:h_cas}
    \mathcal{H}_\mathrm{CAS}
    = \sum_{ij \subset \mathrm{CAS}} (h_{ij} + V^\mathrm{core}_{ij}) c^\dagger_i c_j \\
    + \frac{1}{2} \sum_{ijkl \subset \mathrm{CAS}} g_{ijkl} c^\dagger_i c^\dagger_j c_l c_k
    ,
\end{multline}
where $V^\mathrm{core}$ describes the Coulomb repulsion and exchange interaction with the electrons of the core space, in a partitioning which is common in the field of quantum chemistry \cite{Roos1980,olsen11}.
The CAS Green's~function can then be calculated using the correction vector method described in Sec.~\ref{sec:corrvec}, but
with the reduced dimensionality subspace Hamiltonian $\mathcal{H}_\mathrm{CAS}$ of Eq.~\eqref{eq:h_cas} used in place of~$\mathcal{H}$.

To go from the active space Green's~function to a Green's~function of the entire system, the core part has to be accounted for.
The `formally correct' approach to this involves an inversion of the resulting $G_\mathrm{CAS}$, to calculate an active space self-energy according to the Dyson equation, as
\begin{equation}\label{eq:s_cas}
    \Sigma_{\mathrm{CAS}}(z) = G_{0,\mathrm{CAS}}^{-1}(z) - G_{\mathrm{CAS}}^{-1}(z) ,
\end{equation}
where~$G_{0,\mathrm{CAS}}$ is the non-interacting active space Green's~function \cite{PhysRevB.101.045126}.
The self-energy of Eq.~\eqref{eq:s_cas} can then be added to the (generalized) Fock matrix, $F$, of the CAS $\oplus$ core system, thus defining the full system Green's~function
\begin{equation}\label{eq:g_casci_dyson}
G(z) = \frac{1}{z - F - \Sigma_\mathrm{CAS}(z)}
.
\end{equation}

However, while this is formally correct (and similar to a self-consistent step in dynamical mean-field theory \cite{Bauer2016}), there are some considerations to take into account in this context. The quantum-derived Green's function in the CAS space will contain random noise, from the sampling of the required expectation values such as those in Eqs.~\ref{eq:expect_pauli_ops} and \ref{eq:def_overlapterm}. The result is that this random error will manifest as systematic error in the self-energy due to the non-linear inversion operations required in Eqs.~\ref{eq:s_cas} and \ref{eq:g_casci_dyson} \cite{PhysRevB.98.085118}.
For this reason, we also consider a simpler, but approximate scheme. In this, the non-interacting Green's function~$G_0(z) = \left[ z - F \right]^{-1}$, and the CAS Green's function are combined directly,
i.e.
\begin{equation}\label{eq:g_casci_nondyson}
G(z) = G_0(z) + P_\mathrm{CAS} \left[ G_\mathrm{CAS}(z) - G_0(z) \right] P_\mathrm{CAS}
,
\end{equation}
where $P_\mathrm{CAS}$ is the projector onto the CAS space.
The two approaches become identical, if all of the off-diagonal elements of the generalized Fock matrix~$F$ which couple the CAS and core spaces are zero. In this case, the operations of matrix inversion and projection commute and the approaches should be identical.
While this is not generally the case, $F$ is often still block-diagonally dominant, and therefore the Green's~function of Eq.~\eqref{eq:g_casci_nondyson}
becomes a good approximation to Eq.~\eqref{eq:g_casci_dyson}. Importantly however, this avoids the necessity of matrix inversions, which would themselves introduce a systematic error when performed on stochastically sampled active space Green's functions due to the matrix inversions.
In Sec.~\ref{sec:numerical_results}, we will investigate the use of these active space approaches to Greens functions, and compare the Green's function calculated from Eq.~\ref{eq:g_casci_dyson}, which we will term `Dyson-CAS', with that of the approximation of Eq.~\ref{eq:g_casci_nondyson} (`Non-Dyson-CAS'), in order to quantify this approximation which avoids matrix inversion.

\section{Numerical results\label{sec:numerical_results}}

In this section, we present the results of quantum simulation experiments to test the quantum correction vector VQE on two different molecules, hydrogen (\ce{H_2}) and lithium hydride (\ce{LiH}). Their Hamiltonians are obtained using the STO-3G basis. For the \ce{LiH} molecule, we have further tested the embedding of an active space sampling of the correlated Greens function in a low-energy subspace, detailed in Sec.~\ref{sec:qchem-cas}, to reduce the required number of active spin-orbitals to four. The Hamiltonian is then mapped to a weighted sum of Pauli operators through the Jordan--Wigner mapping \citep{Jordan1928}.

We focus our attention on the accuracy of the single particle spectrum of the systems, for which we define $G(z)$ as the trace of the imaginary part of the full Green's function matrix on either the real or Matsubara axis, as
\begin{equation}
    G(z) = \sum_{\alpha} \mathrm{Im} \left[G_{\alpha \alpha}(z)\right].
\end{equation}
Furthermore, we restrict our simulations to conserve spin symmetry, by working in a restricted formalism where the two spin channels are constrained to be the same, precluding single-particle spin symmetry breaking.
The simulation is realized using the computing platform Hyrax developed by Rahko, with the matrix elements defining the molecular Hamiltonian of each system computed using the {\tt PySCF} simulation package \cite{pyscf1,pyscf2}.

\subsection{Noise model}

In order to model the performance of realistic quantum devices, we include the effects of noises coming from both the variance in the measurements, as well as a two-qubit depolarization noise arising from simulated decoherence of the qubits due to environmental interactions. In particular, this noise model in the two-qubit depolarizing channel is defined as
\begin{align}
    \label{eq:2qb_dep}
    \mathcal{N}(\rho) = 
    (1-p_2) \rho+\frac{p_2}{15}\left(\sum_{a,b} \sigma_{a}^{(1)} \sigma_{b}^{(2)} \rho \sigma_{a}^{(1)} \sigma_{b}^{(2)}\right)-\frac{p_2}{15} \rho,
\end{align}
where $\sigma^{(1)}$ ($\sigma^{(2)}$) denotes a Pauli matrix operating on the first (second) qubit that each two-qubit gate acts on, and $a,b$ sums over all possible Pauli matrices (including the identity) \cite{nielsen_chuang_2010}. This two-qubit depolarizing noise model is applied after every two-qubit gate in the circuit. The $p_2$ parameter is set to $10^{-3}$ for our numerical experiments in Sec~\ref{sec:numerical_results}. On current generation quantum devices, the one-qubit error rate is at least one order of magnitude smaller than the two-qubit error rate and therefore we assume that the one-qubit errors are negligible in the experiments. Furthermore, due to the large number of two-qubit gates in the circuit of Fig.~\ref{fig:overlap circuit}, we anticipate two-qubit errors to be the dominant errors in the computation. We note that when three-qubit gates CCX and CCZ are used for the controlled unitary gates in Fig.~\ref{fig:overlap circuit}, they are decomposed into one and two-qubit gates as done in the work of \citet{on_cnot_cost}.


In order to mitigate the effect of this two-qubit noise, we use a simple exponential extrapolation scheme in our sampling of expectation values, as introduced in Ref.~\onlinecite{EndoExp}, and realized experimentally in Ref.~\onlinecite{DigitalZNE}. To do this, we first execute the circuit with the original noise strength of $p_2=10^{-3}$, and subsequently boost the noise in the simulation by setting $p_2$ to twice its original value. On a physical quantum device this boosting of noise in particular channels can be achieved in several different approaches \cite{Cincio2018,DigitalZNE}. The expectation values of the measurements at these different noise levels are fit to an exponential form as $E(p_2)=E(0) e^{-b p_2}$ to obtain the extrapolated noise-free estimate of $E(0)$. Finally, to converge the effect of measurement noise due to the intrinsic variance in the expectation values, we repeat measurements $10^6$ times to control the statistical accuracy.

\subsection{Simulation details}

\begin{figure}
\begin{center}
\begin{quantikz}
    & \qw & \gategroup[wires=3,steps=4,style={inner sep=6pt}]{Repeated $d_{\mathrm{HEA}}$ times} \qw & \gate{S} & \ctrl{1} & \qw & \gate{S} & \qw  \\
    & \qw & \qw & \gate{S} & \targ{} & \ctrl{1}  & \gate{S} & \qw \\
    & \qw & \qw & \gate{S} & \qw & \targ & \qw & \gate{S} & \qw
\end{quantikz}
\caption{An illustration of the hardware efficient ansatz for three qubits. 
Each $S$ represents a series of single qubit rotation
gates. Each $S$ gate maintains its own angle, and can be adjusted
independently. The depth of the ansatz is denoted by $d_\mathrm{HEA}$.}
\label{fig:HEA_ansatz}
\end{center}
\end{figure}
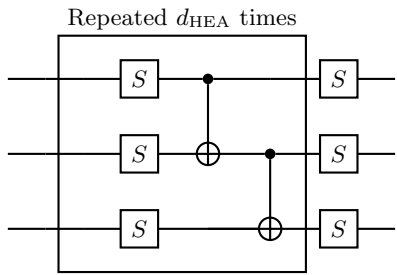

For both the ground state and correction vector circuits, we used the Hardware Efficient Ansatz (HEA) \cite{HEAAnsatz}, illustrated in Fig.~\ref{fig:HEA_ansatz}. The single qubit gates ($S$ in Fig.~\ref{fig:HEA_ansatz}) and number of repeated blocks (depth $d_\mathrm{HEA}$ in Fig.~\ref{fig:HEA_ansatz}) were dynamically adjusted to converge the results without overparameterization of the circuit. The angles, $\theta$, in these single-qubit gates were adjusted to minimize the cost function for the ground state and correction vector circuits (Eq.~\ref{eq:g}) using the Rotosolve algorithm \citep{Rotosolve}. 
We note that for the optimization of the correction vector, the cost function $g(\theta;z,j)$ has a theoretical minimum of $0$, and therefore the optimization is continued until a small
threshold value of $\varepsilon > 0$ is achieved. Typically, we use $\varepsilon=0.05$, and decrease this value if the
obtained Green's function at specific frequency points deviates significantly from its values at neighbouring frequencies. Furthermore, the circuit depths for the correction vectors, $d_{\mathrm{HEA}}$ can vary for different values of $z$. If significant residual values of $g(\theta;z,j)$ remain, then the circuit depth can be increased to provide a more expressive ansatz for the correction vector. 

These variations in the precision required for the optimization and flexibility of the circuit as $z$ is changed are closely tied to the condition number of the operator~$H'(z,j)$. Large condition numbers are found near the poles of the Green's function on the real-frequency axis, which is regularized by the choice of $\eta$, and therefore convergence is faster and requires shallower circuits away from these poles, or on the Matsubara frequency axis. To accelerate the optimization, we initialize the circuit angles for the correction vector optimization by reusing optimized angles from nearby frequencies for subsequent optimization points.

\subsection{Hydrogen molecule}

\begin{figure*}
    \centering
    \subfloat[Retarded Green's function at $d=2\text{\;\r{A}}$]{%
        \label{fig:h2-d2-real}%
        \includegraphics[width=0.45\linewidth]{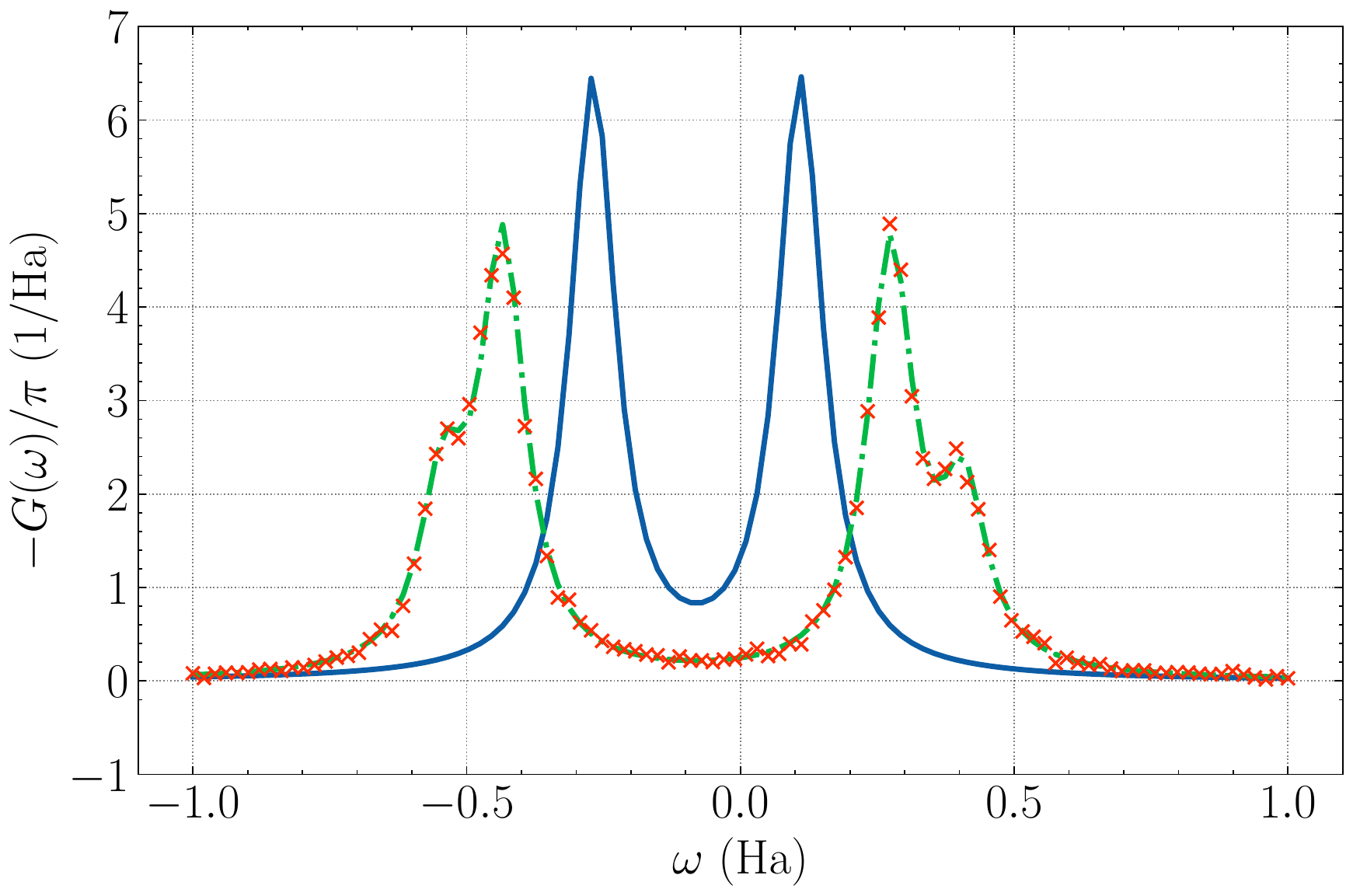}
    }%
    \hspace{8pt}%
    \subfloat[Matsubara Green's function at $d=2\text{\;\r{A}}$]{%
        \label{fig:h2-d2-imag}%
        \includegraphics[width=0.45\linewidth]{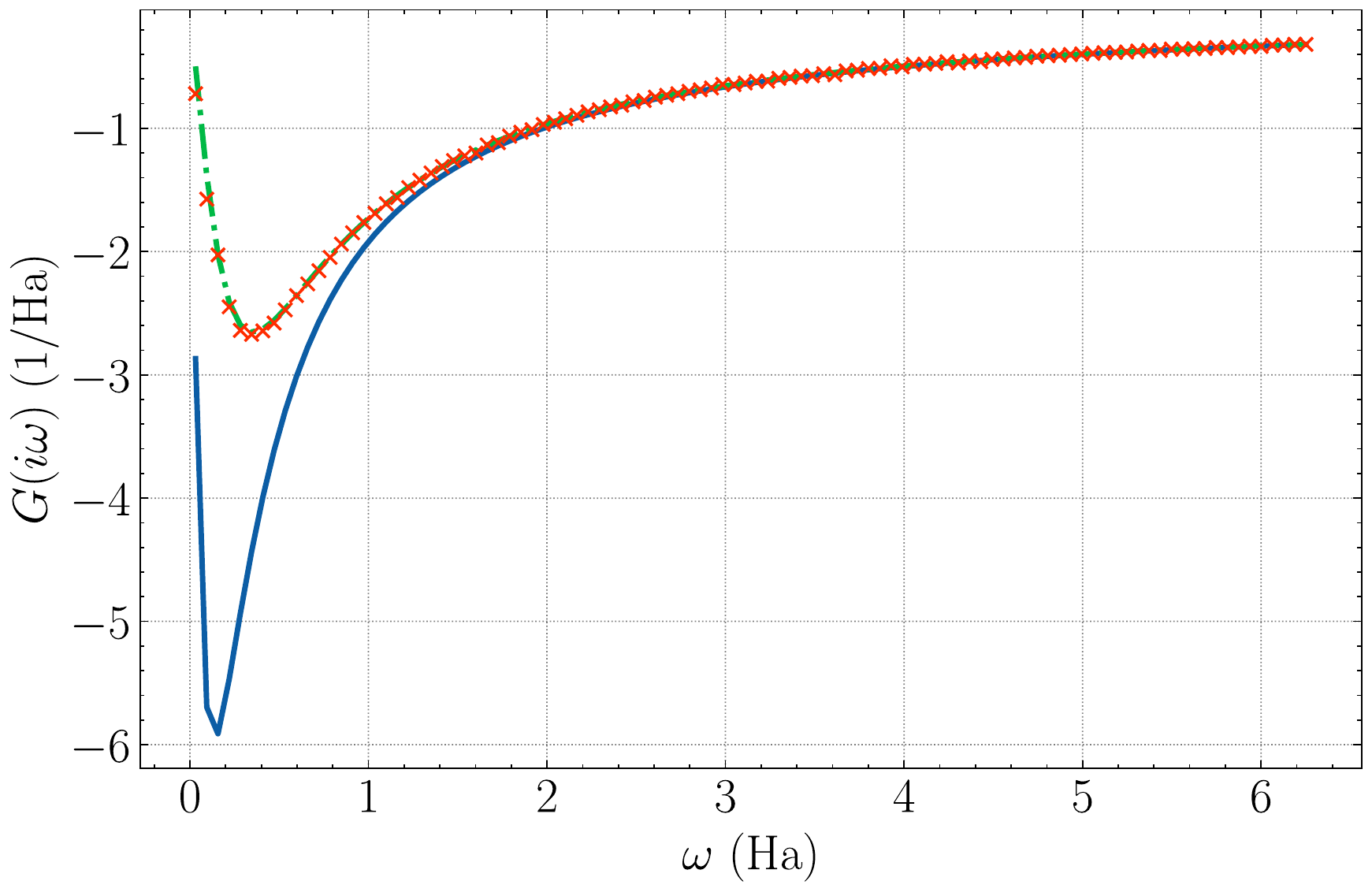}
    }%
    \\
    \subfloat[Retarded Green's function at $d=3\text{\;\r{A}}$]{%
        \label{fig:h2-d3-real}%
        \includegraphics[width=0.45\linewidth]{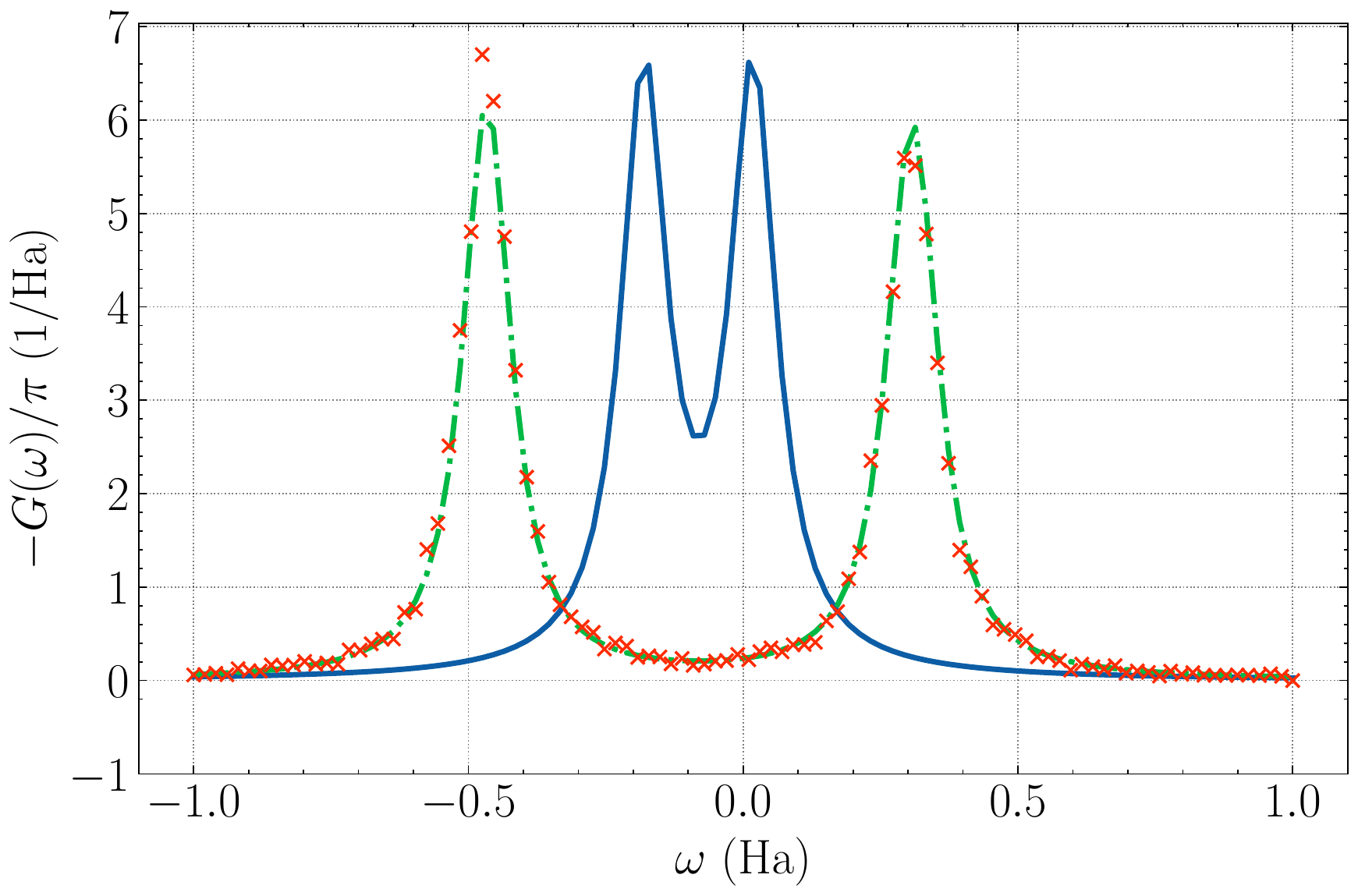}
    }%
    \hspace{8pt}%
    \subfloat[Matsubara Green's function at $d=3\text{\;\r{A}}$]{%
        \label{fig:h2-d3-imag}%
        \includegraphics[width=0.45\linewidth]{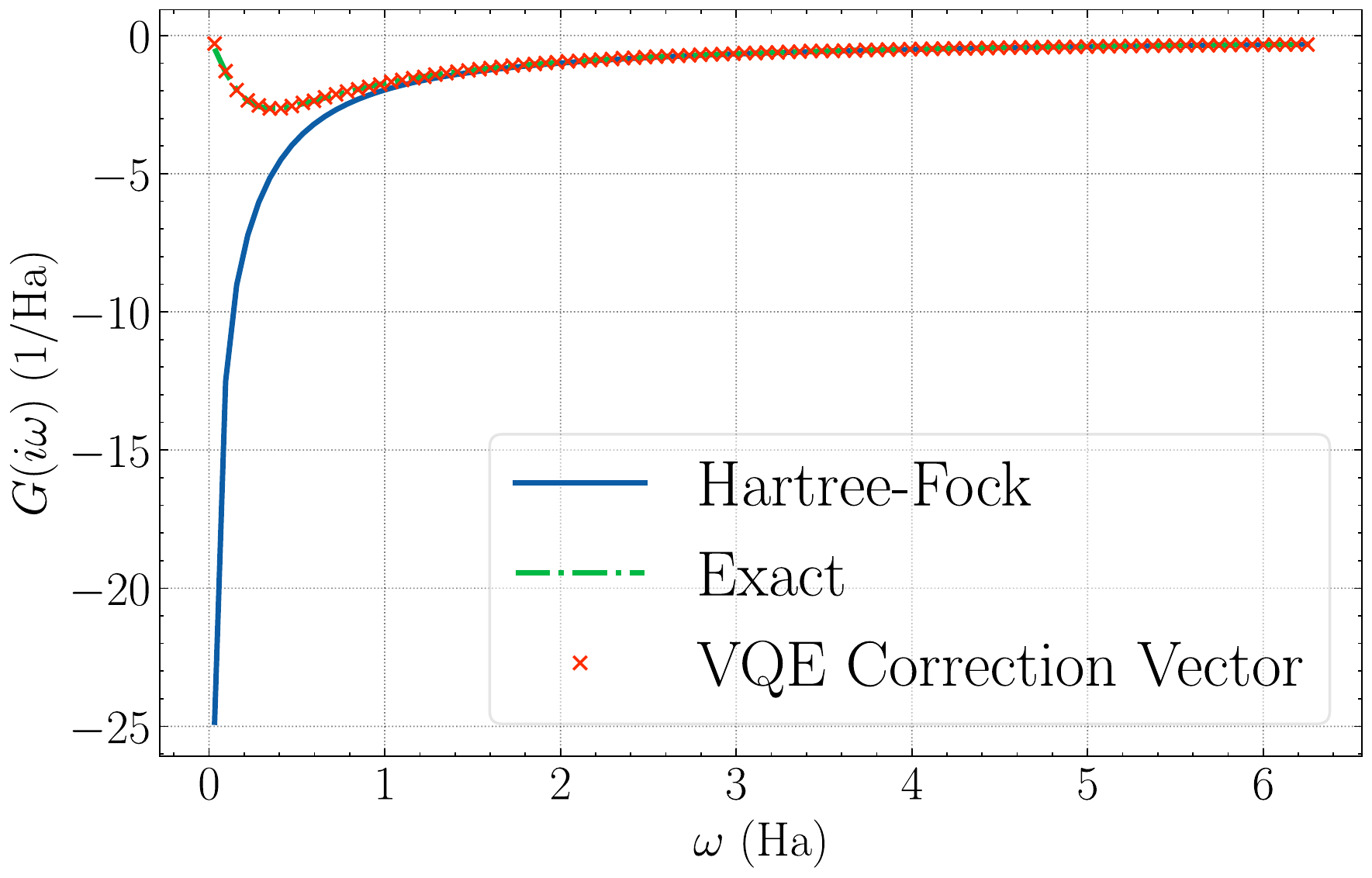}
    }%
    \caption[Matsubara Green's function and retarded Green's function calculated for \ce{H_2} at different distance $d$]{
    Single-particle spectrum of \ce{H_2} computed at two different bond lengths, on both the real-frequency and Matsubara axis. This spectrum is computed from the trace of the imaginary part of the retarded Green's function, calculated via a quantum simulation of the VQE correction vector algorithm  (VQE Correction Vector) and compared to
        the exact solution and uncorrelated mean-field limit (Hartree--Fock).
    }
    \label{fig:H2_GF_plots}%
\end{figure*}

Figure~\ref{fig:H2_GF_plots} shows the simulated single-particle spectrum for the \ce{H_2} molecule at two different bond lengths. These spectra are computed on both the real-frequency (where a broadening of $\eta=0.05$\;Ha is used) and Matsubara axis. In both of these domains the effect of the correlated physics of the system on the single-particle spectrum is clear, with the interacting physics opening the fundamental gap of the system. We expect the correlation effects and entanglement of the orbitals to be larger at the stretched geometry of 3\;\AA, where this opening of the charge gap is most significant and where the situation is similar to the opening of a Mott gap due to interactions from a condensed matter perspective. At less stretched geometries, correlation-induced substructures are observed at this energy resolution, splitting the peaks compared to the mean-field solution at this basis size.

The quantum correction vector approach reproduces the exact spectrum to high fidelity, where the exact results are computed via an exact inversion of the full Hamiltonian on classical computers for these small system sizes. The discrepancy on the Matsubara axis is almost unobservable on the scale of the plot, and only has a maximum of $0.22\;\text{Ha}^{-1}$ over the entire domain of the Matsubara axis. Note that the analytic properties of the Green's function ensure that the same information is contained in both the Matsubara and real-frequency domains. On the real-frequency axis, the discrepancies primarily manifest as noise in the transition amplitudes at the excitation energies. All of these observations are consistent with the condition number of $H'(z,j)$ being the determining factor in the accuracy of the method, and hence why an adaptive approach to ansatz expressibility and convergence thresholds is useful in these regimes. Furthermore, we should note that formal properties such as sum rules and causality which should be present in the final Green's function are obeyed to good accuracy, and at full convergence should be obeyed exactly.

\subsection{Lithium Hydride}

\begin{figure}
    \centering
    \subfloat[$d=2$, Retarded Green's function]{%
        \label{fig:lih-d2-real}%
        \includegraphics[width=0.45\textwidth]{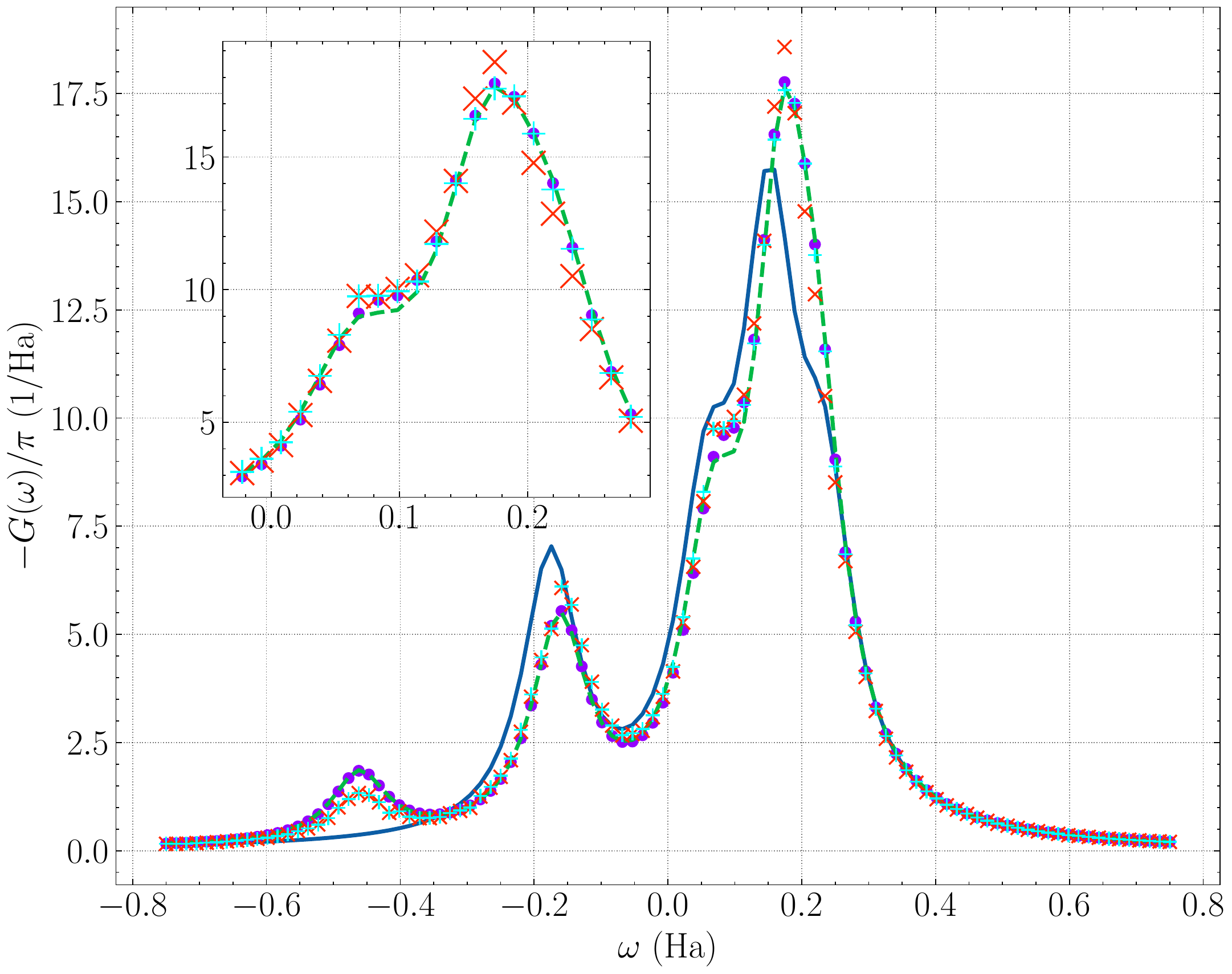}
    }%
    \hspace{8pt}%
    \subfloat[$d=2$, Matsubara Green's function]{%
        \label{fig:lih-d2-imag}%
        \includegraphics[width=0.45\textwidth]{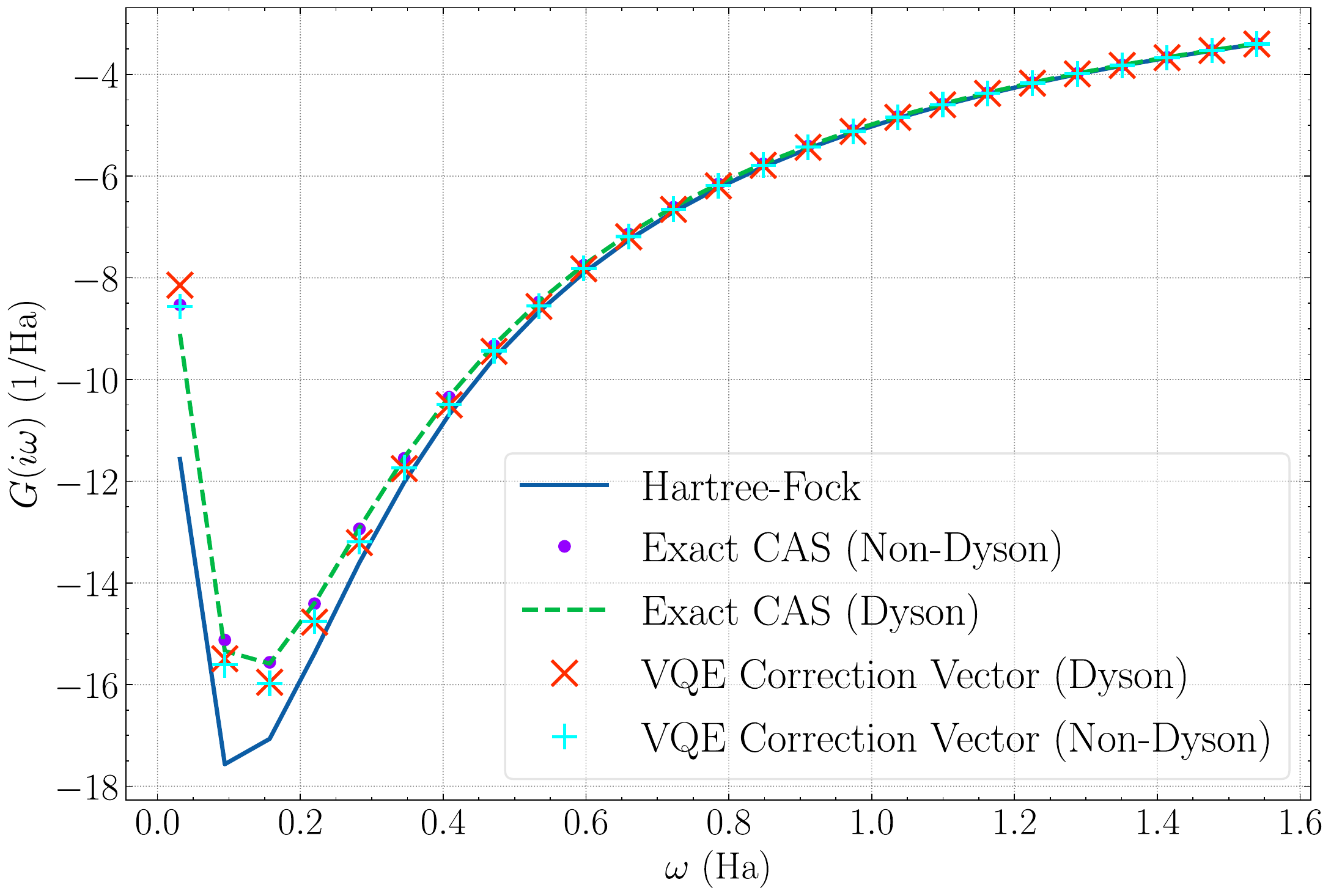}
    }%
    \caption[Matsubara Green's function and retarded Green's function calculated for the lithium hydride molecule at distance $d=3$ angstrom]{
        Single-particle spectral function for \ce{LiH} on both the real-frequency and Matsubara axis. Results are presented from the embedding of a two-orbital, low-energy active space, in the deep-lying core $\sigma$ orbital at the mean-field level. Results are shown from quantum simulation with the VQE correction vector `VQE Correction Vector (Non-Dyson)', exact classical approaches (both with `Dyson' embedding and `Non-Dyson' embedding), and the uncorrelated mean-field result (`Hartree--Fock'). The inset in Fig.4.a emphasizes the region of the retarded Green's function where there is a noticeable difference between the data obtained from VQE correction vector method in the two different approaches.
    }
    \label{fig:LiH}%
\end{figure}

In Figure~\ref{fig:LiH} we apply the method to the \ce{LiH} molecule, at a slightly stretched bond length of 2\;\AA.  For this system we take an active space of the highest energy occupied and lowest energy unoccupied degrees of freedom from a prior Hartree--Fock calculation as a four-qubit space in which to solve for the correlated VQE correction vector. This leaves two electrons in a low-energy $\sigma$-bonding orbital which are considered inert for the purposes of the entanglement and correlation effects with the rest of the electrons, and therefore just contribute a Coulomb and exchange static potential to the active space Hamiltonian, as described in Sec.~\ref{sec:qchem-cas}. The correlation is found to induce a satellite peak at energy ranges significantly outside the energy window of the orbitals chosen. 

There are two approaches to combine the response of this inactive core orbital with the rest of the system, which were termed `Dyson' and `Non-Dyson' in Sec.~\ref{sec:qchem-cas}. The former approach combines the self-energies in the different spaces and is formally correct. However, it requires inversion of Green's function estimates from the quantum solver, which will introduce systematic error in the final result for any non-zero random error. In contrast, the `Non-Dyson' approach combines the Greens functions directly, neglecting the effect of the off-diagonal block of the generalized Fock matrix coupling the spaces in the full system mean-field spectrum. However, this does not require any inversion of the sampled Green's function, and since the initial orbitals are designed to have no coupling in the mean-field picture, it is only the presence of correlations which can subsequently induce a small coupling between the active space and core orbital. Furthermore, from a computational perspective, the `Dyson' embedding requires the calculation of the off-diagonal Green's function elements in the active space, however this does not require the optimization of any further correction vectors.

The assumption that the difference between the `Dyson' and approximate `non-Dyson` embedding is slight is corroborated by the results of Fig.~\ref{fig:LiH}, where we can see that for this system the two approaches for an exact active space Green's function embedded in the wider system are virtually indistinguishable. This is expected for this choice of active space embedding, where the correlated subspace is chosen from canonical Hartree--Fock orbitals and where the hybridization in the final state between the active space and core orbital is expected to be small. From another perspective, the differences are expected to be minor in the case where there is not significant relaxation of the full system electronic density due to the correlation effects. 

We can also observe the differences between these approaches when the active space Green's function is solved with the quantum simulator and significant simulated quantum noise. Both of the quantum-simulated VQE solutions to the embedded active space Green's function again reproduce the correlated spectrum to a high degree of accuracy, with only relatively minor differences arising from the magnitude of the transition amplitudes at the excitation energies. However, close to the poles, the accuracy of the `Non-Dyson' approach is marginally superior, indicating that the bias due to the inversion of the noisy Green's function matrix is more significant than the neglect of the hybridization between the active space and core orbital for this particular choice of active space orbitals. While this is an interesting observation, this conclusion is likely to still be sensitive to the specific choice of active space orbitals, molecular system, and noise in the sampled active space Green's function. The low-energy satellite peak in Fig.~\ref{fig:LiH} at $\sim-0.45$~Ha is however underestimated slightly in both quantum approaches.

\subsection{Perspective}\label{sec:discussion}

It is important to consider the scalability and applicability of the VQE correction vector algorithm to current and near-term
quantum devices \citep{Preskill2018}.
The algorithm prepares both the ground state and the correction vector for different frequency points
using circuits representing a variational ansatz. The approach of parameterized quantum circuits has been widely used for the practical simulation of both ground and excited states in recent years \cite{mcardleQuantumComputationalChemistry2018, McClean2017, Colless2018, Santagati2018, Higgott2019, Jones2019, McArdle2019, Ollitrault2020, Tilly2020}, although the scalability of the optimization problem for the gate parameters is a source of debate in the literature \cite{PhysRevB.102.075104,wang2021noiseinduced,Grant2019initialization}. This is expected to be compounded by the condition number of the~$H'(z,j)$ operator generally being larger than that of its parent Hamiltonian used to simulate the ground state.
The correction vector~$|\chi_j(z)\rangle$ at each frequency contains contributions from all the excited states of $\mathcal{H}$ in the probed symmetry sector. However, on the real-frequency axis, this combination of states will be dominated by the excited state corresponding to the nearest excitation energy to the sampled frequency.
In this case, we expect the required ansatz complexity (depth, for example) to express the correction vector to be similar to the required ansatz complexity to describe these excited states, which will again depend on the level of entanglement between the degrees of freedom.
Therefore, we expect the scalability of our correction vector ansatz to be comparable
to other state-specific VQE-based algorithms \cite{DMFTIvan,tilly2021reduced}.

We can also consider the overall circuit complexity in this approach compared to ground-state VQE implementations. This complexity is increased due to the requirement to sample overlaps of the type described in Sec.~\ref{sec:implementation}, which require a significant number of controlled-unitary gates. This has inhibited experiments on the current generation of quantum computers, due to their relatively high two-qubit gate error rates. However, the largest circuit depths required in the correction vector VQE algorithm (in particular, required for the calculation of $g(\theta)$ in Eq.~\ref{eq:g}) is only twice that of the circuit depths of a ground state VQE algorithm. Furthermore, only a single ancillary qubit is required for the measurements of overlaps. This certainly puts it in the range of near-term devices when the bottleneck of the two-qubit gate fidelity is improved with respect to noise. Furthermore, all optimization algorithms designed to perform the classical parameter updates can be straightforwardly ported from ground state VQE codes \cite{Rotosolve}, as well as potentially more sophisticated approaches such as the AdaptVQE for ansatz selection \cite{Adaptvqe}.


\section{Conclusions}

In this paper, we propose an approach for the simulation of arbitrary dynamic correlation functions on quantum devices, solving directly in the (real or Matsubara) frequency domain via a coupled quantum-classical variational approach.
The method combines the conventional VQE and a variational linear equation solver to prepare the correction vector via a parameterized circuit ansatz in order to calculate the single-particle Green's function of correlated molecular systems.
The approach can converge to the exact Green's function, assuming a sufficient expressibility in the circuit representing the ground and linear response states, with a circuit depth which is then independent of the resolution on the frequency axis. This contrasts with state-specific approaches to constructing Green's functions which start from the spectral representation, and which suffer from a cutoff in the energy resolution \cite{DMFTIvan,Endo2019}, or time-evolution methods where the length of time the state is propagated for (which generally changes the circuit depth) controls the frequency resolution of the resulting dynamical function \cite{Bauer2016, Kreula2016, Kreula2016_, Keen2020}. Although the overall gate depths are only a factor of two longer than a ground-state VQE approach, the increase in the number of two-qubit entangling gates results in the algorithm being particularly sensitive to the levels of two-qubit gate noise.

In our numerical simulations, the method reproduced the Green's function over both real and imaginary frequencies for the \ce{H_2} and \ce{LiH} molecules to high accuracy, under an simulated noise model for errors in the dominating two-qubit gates. We focus both on simulation of the whole Green's function, and also on an approach to isolate a correlated low-energy `active' subspace in which to simulate the correlated dynamics. These dynamics can be embedded into the mean-field spectrum of the whole system either via a combination of Green's functions, or self-energies. While the latter approach is formally correct, it requires inversion of the noisy sampled Green's function matrix. We demonstrate that the Green's function approach to combining the response of the active and external spaces is sufficiently accurate for the \ce{LiH} system, where the correlation effects are not too strong, and the hybridization between these canonical degrees of freedom is small. Future work will look to expand this approach to other dynamical correlation functions of interest, as well as investigate the use of the approach within a self-consistent dynamical mean-field embedding theory \cite{Bauer2016,tilly2021reduced,Keen2020,DMFTIvan,GalliQuantumEmbedding} and assess the viability of the approach on physical quantum devices.

\begin{acknowledgments}

H.C. is supported through a Teaching Fellowship from UCL.
J.T. is supported by and industrial CASE (iCASE) studentship, funded by and UK EPSRC [EP/R513143/1], in collaboration with University College London and Rahko Ltd. G.H.B. gratefully acknowledges support from the Royal Society via a University Research Fellowship, as well as funding from the European Research Council (ERC) under the European Union\textquotesingle s Horizon 2020 research and innovation programme (Grant Agreement No. 759063). This research project is in great part funded by Innovate UK, project Quantifi from UK Research and Innovation, as part of the UK National Quantum Technologies Programme and Industrial Strategy Challenge Fund. We thank the members of the Amazon Quantum Solutions Lab at Amazon Web Services (AWS) for their valued support of project Quantifi, and provision of computational resources used in this work. The views expressed are those of the authors and do not reflect the official policy or position of AWS or the members of the Amazon Quantum Solutions Lab.

\end{acknowledgments}


\bibliographystyle{unsrt}

\begin{thebibliography}{10}

\bibitem{Coey2010}
J.~M.~D. Coey.
\newblock {\em Magnetism and Magnetic Materials}.
\newblock Cambridge University Press, January 2001.

\bibitem{Hasan2010}
M.~Zahid Hasan and Charles~L. Kane.
\newblock Colloquium: Topological insulators.
\newblock {\em Reviews of Modern Physics}, 82(4):3045--3067, November 2010.

\bibitem{Imada1998}
Masatoshi Imada, Atsushi Fujimori, and Yoshinori Tokura.
\newblock Metal-insulator transitions.
\newblock {\em Rev. Mod. Phys.}, 70:1039--1263, Oct 1998.

\bibitem{Bednorz1986}
J.~G. Bednorz and K.~A. M\"{o}ller.
\newblock Possible {highT} c superconductivity in the ba-la-cu-o system.
\newblock {\em Zeitschrift f\"{u}r Physik B Condensed Matter}, 64(2):189--193,
  June 1986.

\bibitem{Droghetti2017}
Andrea Droghetti and Ivan Rungger.
\newblock Quantum transport simulation scheme including strong correlations and
  its application to organic radicals adsorbed on gold.
\newblock {\em Phys. Rev. B}, 95:085131, Feb 2017.

\bibitem{Kovaleva2008}
Elena~G Kovaleva and John~D Lipscomb.
\newblock Versatility of biological non-heme fe({II}) centers in oxygen
  activation reactions.
\newblock {\em Nature Chemical Biology}, 4(3):186--193, February 2008.

\bibitem{Weber2014}
C.~Weber, D.~J. Cole, D.~D. O{\textquotesingle}Regan, and M.~C. Payne.
\newblock Renormalization of myoglobin-ligand binding energetics by quantum
  many-body effects.
\newblock {\em Proceedings of the National Academy of Sciences},
  111(16):5790--5795, April 2014.

\bibitem{Lloyd1073}
Seth Lloyd.
\newblock Universal quantum simulators.
\newblock {\em Science}, 273(5278):1073--1078, 1996.

\bibitem{doi:10.1080/00268976.2011.552441}
James~D. Whitfield, Jacob Biamonte, and Alain Aspuru-Guzik.
\newblock Simulation of electronic structure hamiltonians using quantum
  computers.
\newblock {\em Molecular Physics}, 109(5):735--750, 2011.

\bibitem{Cao2019}
Yudong Cao, Jonathan Romero, Jonathan~P. Olson, Matthias Degroote, Peter~D.
  Johnson, M{\'{a}}ria Kieferov{\'{a}}, Ian~D. Kivlichan, Tim Menke, Borja
  Peropadre, Nicolas P.~D. Sawaya, Sukin Sim, Libor Veis, and Al{\'{a}}n
  Aspuru-Guzik.
\newblock Quantum chemistry in the age of quantum computing.
\newblock {\em Chemical Reviews}, 119(19):10856--10915, August 2019.

\bibitem{Bauer2020}
Bela Bauer, Sergey Bravyi, Mario Motta, and Garnet Kin-Lic Chan.
\newblock Quantum algorithms for quantum chemistry and quantum materials
  science.
\newblock {\em Chemical Reviews}, 120(22):12685--12717, October 2020.

\bibitem{mcardleQuantumComputationalChemistry2018}
Sam McArdle, Suguru Endo, Al{\'{a}}n Aspuru-Guzik, Simon~C. Benjamin, and Xiao
  Yuan.
\newblock Quantum computational chemistry.
\newblock {\em Reviews of Modern Physics}, 92(1), March 2020.

\bibitem{Cleve1998}
R.~Cleve, A.~Ekert, C.~Macchiavello, and M.~Mosca.
\newblock Quantum algorithms revisited.
\newblock {\em Proceedings of the Royal Society of London. Series A:
  Mathematical, Physical and Engineering Sciences}, 454(1969):339--354, January
  1998.

\bibitem{Kitaev1996}
A.~Yu. Kitaev.
\newblock Quantum measurements and the abelian stabilizer problem.
\newblock {\em arXiv:quant-ph/9511026}, 1995.

\bibitem{Abrams1999}
Daniel~S. Abrams and Seth Lloyd.
\newblock Quantum algorithm providing exponential speed increase for finding
  eigenvalues and eigenvectors.
\newblock {\em Physical Review Letters}, 83(24):5162--5165, December 1999.

\bibitem{Preskill2018}
John Preskill.
\newblock Quantum {C}omputing in the {NISQ} era and beyond.
\newblock {\em {Quantum}}, 2:79, August 2018.

\bibitem{VQE}
Alberto Peruzzo, Jarrod McClean, Peter Shadbolt, Man-Hong Yung, Xiao-Qi Zhou,
  Peter~J. Love, Al{\'{a}}n Aspuru-Guzik, and Jeremy~L. O'Brien.
\newblock A variational eigenvalue solver on a photonic quantum processor.
\newblock {\em Nature Communications}, 5(1), July 2014.

\bibitem{McClean2017}
Jarrod~R. McClean, Mollie~E. Kimchi-Schwartz, Jonathan Carter, and Wibe~A.
  de~Jong.
\newblock Hybrid quantum-classical hierarchy for mitigation of decoherence and
  determination of excited states.
\newblock {\em Physical Review A}, 95(4), April 2017.

\bibitem{Colless2018}
J.~I. Colless, V.~V. Ramasesh, D.~Dahlen, M.~S. Blok, M.~E. Kimchi-Schwartz,
  J.~R. McClean, J.~Carter, W.~A. de~Jong, and I.~Siddiqi.
\newblock Computation of molecular spectra on a quantum processor with an
  error-resilient algorithm.
\newblock {\em Phys. Rev. X}, 8:011021, Feb 2018.

\bibitem{Santagati2018}
Raffaele Santagati, Jianwei Wang, Antonio~A. Gentile, Stefano Paesani, Nathan
  Wiebe, Jarrod~R. McClean, Sam Morley-Short, Peter~J. Shadbolt, Damien
  Bonneau, Joshua~W. Silverstone, David~P. Tew, Xiaoqi Zhou, Jeremy~L. O'Brien,
  and Mark~G. Thompson.
\newblock Witnessing eigenstates for quantum simulation of hamiltonian spectra.
\newblock {\em Science Advances}, 4(1):eaap9646, January 2018.

\bibitem{Higgott2019}
Oscar Higgott, Daochen Wang, and Stephen Brierley.
\newblock Variational quantum computation of excited states.
\newblock {\em Quantum}, 3:156, July 2019.

\bibitem{Jones2019}
Tyson Jones, Suguru Endo, Sam McArdle, Xiao Yuan, and Simon~C. Benjamin.
\newblock Variational quantum algorithms for discovering hamiltonian spectra.
\newblock {\em Physical Review A}, 99(6), June 2019.

\bibitem{McArdle2019}
Sam McArdle, Tyson Jones, Suguru Endo, Ying Li, Simon~C. Benjamin, and Xiao
  Yuan.
\newblock Variational ansatz-based quantum simulation of imaginary time
  evolution.
\newblock {\em npj Quantum Information}, 5(1), September 2019.

\bibitem{Ollitrault2020}
Pauline~J. Ollitrault, Abhinav Kandala, Chun-Fu Chen, Panagiotis~Kl.
  Barkoutsos, Antonio Mezzacapo, Marco Pistoia, Sarah Sheldon, Stefan Woerner,
  Jay~M. Gambetta, and Ivano Tavernelli.
\newblock Quantum equation of motion for computing molecular excitation
  energies on a noisy quantum processor.
\newblock {\em Physical Review Research}, 2(4), October 2020.

\bibitem{Tilly2020}
Jules Tilly, Glenn Jones, Hongxiang Chen, Leonard Wossnig, and Edward Grant.
\newblock Computation of molecular excited states on {IBM} quantum computers
  using a discriminative variational quantum eigensolver.
\newblock {\em Physical Review A}, 102(6), December 2020.

\bibitem{zhangAdaptiveVariationalQuantum2021}
Feng Zhang, Niladri Gomes, Yongxin Yao, Peter~P. Orth, and Thomas Iadecola.
\newblock Adaptive variational quantum eigensolvers for highly excited states.
\newblock {\em arXiv:2104.12636}, April 2021.

\bibitem{ML_Corrvec}
Douglas Hendry and Adrian~E. Feiguin.
\newblock Machine learning approach to dynamical properties of quantum
  many-body systems.
\newblock {\em Physical Review B}, 100(24), December 2019.

\bibitem{DMRG_Corrvec}
Till~D. K\"{u}hner and Steven~R. White.
\newblock Dynamical correlation functions using the density matrix
  renormalization group.
\newblock {\em Physical Review B}, 60(1):335--343, July 1999.

\bibitem{DDMRG}
Eric Jeckelmann.
\newblock Dynamical density-matrix renormalization-group method.
\newblock {\em Physical Review B}, 66(4), July 2002.

\bibitem{doi:10.1021/acs.jctc.8b00454}
Pradipta~Kumar Samanta, Nick~S. Blunt, and George~H. Booth.
\newblock Response formalism within full configuration interaction quantum
  monte carlo: Static properties and electrical response.
\newblock {\em Journal of Chemical Theory and Computation}, 14(7):3532--3546,
  June 2018.

\bibitem{PhysRevB.101.045126}
Max Nusspickel and George~H. Booth.
\newblock Frequency-dependent and algebraic bath states for a dynamical
  mean-field theory with compact support.
\newblock {\em Phys. Rev. B}, 101:045126, Jan 2020.

\bibitem{PhysRevB.85.205119}
P.~E. Dargel, A.~W\"ollert, A.~Honecker, I.~P. McCulloch, U.~Schollw\"ock, and
  T.~Pruschke.
\newblock Lanczos algorithm with matrix product states for dynamical
  correlation functions.
\newblock {\em Phys. Rev. B}, 85:205119, May 2012.

\bibitem{Wecker2015}
Dave Wecker, Matthew~B. Hastings, Nathan Wiebe, Bryan~K. Clark, Chetan Nayak,
  and Matthias Troyer.
\newblock Solving strongly correlated electron models on a quantum computer.
\newblock {\em Physical Review A}, 92(6), December 2015.

\bibitem{Bauer2016}
Bela Bauer, Dave Wecker, Andrew~J. Millis, Matthew~B. Hastings, and Matthias
  Troyer.
\newblock Hybrid quantum-classical approach to correlated materials.
\newblock {\em Phys. Rev. X}, 6:031045, Sep 2016.

\bibitem{Kreula2016}
Juha~M Kreula, Laura Garc{\'{\i}}a-{\'{A}}lvarez, Lucas Lamata, Stephen~R
  Clark, Enrique Solano, and Dieter Jaksch.
\newblock Few-qubit quantum-classical simulation of strongly correlated lattice
  fermions.
\newblock {\em {EPJ} Quantum Technology}, 3(1), August 2016.

\bibitem{Kreula2016_}
J.~M. Kreula, S.~R. Clark, and D.~Jaksch.
\newblock Non-linear quantum-classical scheme to simulate non-equilibrium
  strongly correlated fermionic many-body dynamics.
\newblock {\em Scientific Reports}, 6(1), September 2016.

\bibitem{Keen2020}
Trevor Keen, Thomas Maier, Steven Johnston, and Pavel Lougovski.
\newblock {Quantum-classical simulation of two-site dynamical mean-field theory
  on noisy quantum hardware}.
\newblock {\em Quantum Science and Technology}, 5(3):1--10, 2020.

\bibitem{Low2019hamiltonian}
Guang~Hao Low and Isaac~L. Chuang.
\newblock Hamiltonian simulation by qubitization.
\newblock {\em Quantum}, 3:163, July 2019.

\bibitem{PhysRevB.103.014301}
Dries Sels and Eugene Demler.
\newblock Quantum generative model for sampling many-body spectral functions.
\newblock {\em Phys. Rev. B}, 103:014301, Jan 2021.

\bibitem{Endo2019}
Suguru Endo, Iori Kurata, and Yuya~O. Nakagawa.
\newblock Calculation of the green's function on near-term quantum computers.
\newblock {\em Phys. Rev. Research}, 2:033281, Aug 2020.

\bibitem{PRXQuantum.2.010317}
Shi-Ning Sun, Mario Motta, Ruslan~N. Tazhigulov, Adrian~T.K. Tan, Garnet
  Kin-Lic Chan, and Austin~J. Minnich.
\newblock Quantum computation of finite-temperature static and dynamical
  properties of spin systems using quantum imaginary time evolution.
\newblock {\em PRX Quantum}, 2:010317, Feb 2021.

\bibitem{PhysRevLett.103.150502}
Aram~W. Harrow, Avinatan Hassidim, and Seth Lloyd.
\newblock Quantum algorithm for linear systems of equations.
\newblock {\em Physical Review Letters}, 103(15), October 2009.

\bibitem{10.1145/3313276.3316366}
Andr\'{a}s Gily\'{e}n, Yuan Su, Guang~Hao Low, and Nathan Wiebe.
\newblock Quantum singular value transformation and beyond: Exponential
  improvements for quantum matrix arithmetics.
\newblock In {\em Proceedings of the 51st Annual ACM SIGACT Symposium on Theory
  of Computing}, STOC 2019, page 193–204, New York, NY, USA, 2019.
  Association for Computing Machinery.

\bibitem{doi:10.1137/16M1087072}
Andrew~M. Childs, Robin Kothari, and Rolando~D. Somma.
\newblock Quantum algorithm for systems of linear equations with exponentially
  improved dependence on precision.
\newblock {\em SIAM Journal on Computing}, 46(6):1920--1950, 2017.

\bibitem{PhysRevLett.122.060504}
Yi\ifmmode \breve{g}\else~\u{g}\fi{}it Suba\ifmmode \mbox{\c{s}}\else
  \c{s}\fi{}\ifmmode \imath \else~\i \fi{}, Rolando~D. Somma, and Davide
  Orsucci.
\newblock Quantum algorithms for systems of linear equations inspired by
  adiabatic quantum computing.
\newblock {\em Phys. Rev. Lett.}, 122:060504, Feb 2019.

\bibitem{LinLin}
Yu~Tong, Dong An, Nathan Wiebe, and Lin Lin.
\newblock Fast inversion, preconditioned quantum linear system solvers, and
  fast evaluation of matrix functions.
\newblock {\em arXiv:2008.13295}, 2020.

\bibitem{Roggero2019}
Alessandro Roggero and Joseph Carlson.
\newblock Dynamic linear response quantum algorithm.
\newblock {\em Phys. Rev. C}, 100:034610, Sep 2019.

\bibitem{Kosugi2020}
Taichi Kosugi and Yu-ichiro Matsushita.
\newblock Construction of green's functions on a quantum computer:
  Quasiparticle spectra of molecules.
\newblock {\em Phys. Rev. A}, 101:012330, Jan 2020.

\bibitem{caiQuantumComputationMolecular2020}
Xiaoxia Cai, Wei-Hai Fang, Heng Fan, and Zhendong Li.
\newblock Quantum computation of molecular response properties.
\newblock {\em Physical Review Research}, 2(3):033324, August 2020.

\bibitem{DMFTIvan}
I.~Rungger, N.~Fitzpatrick, H.~Chen, C.~H. Alderete, H.~Apel, A.~Cowtan,
  A.~Patterson, D.~Munoz Ramo, Y.~Zhu, N.~H. Nguyen, E.~Grant, S.~Chretien,
  L.~Wossnig, N.~M. Linke, and R.~Duncan.
\newblock Dynamical mean field theory algorithm and experiment on quantum
  computers, 2019.

\bibitem{zhuCalculatingGreenFunction2021}
Jie Zhu, Yuya~O. Nakagawa, Chuan-Feng Li, Guang-Can Guo, and Yong-Sheng Zhang.
\newblock Calculating the {{Green}}'s function of two-site {{Fermionic
  Hubbard}} model in a photonic system.
\newblock {\em arXiv:2104.12361}, April 2021.

\bibitem{vqela}
Xiaosi Xu, Jinzhao Sun, Suguru Endo, Ying Li, Simon~C. Benjamin, and Xiao Yuan.
\newblock Variational algorithms for linear algebra, 2019.

\bibitem{bravyi_fermionic_2002}
Sergey~B. Bravyi and Alexei~Yu. Kitaev.
\newblock Fermionic quantum computation.
\newblock {\em Annals of Physics}, 298(1):210--226, May 2002.

\bibitem{Seeley2012}
Jacob~T. Seeley, Martin~J. Richard, and Peter~J. Love.
\newblock The bravyi-kitaev transformation for quantum computation of
  electronic structure.
\newblock {\em The Journal of Chemical Physics}, 137(22):224109, December 2012.

\bibitem{bravyi_tapering_2017}
Sergey Bravyi, Jay~M. Gambetta, Antonio Mezzacapo, and Kristan Temme.
\newblock Tapering off qubits to simulate fermionic hamiltonians.
\newblock {\em arXiv:1701.08213}, 2017.

\bibitem{Setia2019}
Kanav Setia, Sergey Bravyi, Antonio Mezzacapo, and James~D. Whitfield.
\newblock Superfast encodings for fermionic quantum simulation.
\newblock {\em Physical Review Research}, 1(3), October 2019.

\bibitem{Jordan1928}
Pascual Jordan and Eugene Wigner.
\newblock {{\"{U}}ber das Paulische {\"{A}}quivalenzverbot}.
\newblock {\em Z. Physik}, 47:631--651, sep 1928.

\bibitem{QITE}
Mario Motta, Chong Sun, Adrian T.~K. Tan, Matthew~J. O'Rourke, Erika Ye,
  Austin~J. Minnich, Fernando G. S.~L. Brand{\~{a}}o, and Garnet Kin-Lic Chan.
\newblock Determining eigenstates and thermal states on a quantum computer
  using quantum imaginary time evolution.
\newblock {\em Nature Physics}, 16(2):205--210, November 2019.

\bibitem{Wang2019}
Daochen Wang, Oscar Higgott, and Stephen Brierley.
\newblock Accelerated variational quantum eigensolver.
\newblock {\em Physical Review Letters}, 122(14), April 2019.

\bibitem{yeteraydeniz2021benchmarking}
Kübra Yeter-Aydeniz, Bryan~T. Gard, Jacek Jakowski, Swarnadeep Majumder,
  George~S. Barron, George Siopsis, Travis Humble, and Raphael~C. Pooser.
\newblock Benchmarking quantum chemistry computations with variational,
  imaginary time evolution, and krylov space solver algorithms.
\newblock {\em arXiv:2102.05511}, 2021.

\bibitem{tilly2021reduced}
Jules Tilly, P.~V. Sriluckshmy, Akashkumar Patel, Enrico Fontana, Ivan Rungger,
  Edward Grant, Robert Anderson, Jonathan Tennyson, and George~H. Booth.
\newblock Reduced density matrix sampling: Self-consistent embedding and
  multiscale electronic structure on current generation quantum computers,
  2021.

\bibitem{GalliQuantumEmbedding}
He~Ma, Marco Govoni, and Giulia Galli.
\newblock Quantum simulations of materials on near-term quantum computers.
\newblock {\em npj Computational Materials}, 6(1), July 2020.

\bibitem{yamazaki2018practical}
Takeshi Yamazaki, Shunji Matsuura, Ali Narimani, Anushervon Saidmuradov, and
  Arman Zaribafiyan.
\newblock Towards the practical application of near-term quantum computers in
  quantum chemistry simulations: A problem decomposition approach.
\newblock {\em arXiv:1806.01305}, 2018.

\bibitem{dhawan2021dynamical}
Diksha Dhawan, Mekena Metcalf, and Dominika Zgid.
\newblock Dynamical self-energy mapping (dsem) for quantum computing.
\newblock {\em arXiv:2010.05441}, 2020.

\bibitem{Takeshita2020}
Tyler Takeshita, Nicholas~C. Rubin, Zhang Jiang, Eunseok Lee, Ryan Babbush, and
  Jarrod~R. McClean.
\newblock Increasing the representation accuracy of quantum simulations of
  chemistry without extra quantum resources.
\newblock {\em Phys. Rev. X}, 10:011004, Jan 2020.

\bibitem{Yao2020}
Yongxin Yao, Feng Zhang, Cai-Zhuang Wang, Kai-Ming Ho, and Peter~P. Orth.
\newblock Gutzwiller hybrid quantum-classical computing approach for correlated
  materials.
\newblock {\em Physical Review Research}, 3(1), February 2021.

\bibitem{rossmannek2020quantum}
Max Rossmannek, Panagiotis~Kl. Barkoutsos, Pauline~J. Ollitrault, and Ivano
  Tavernelli.
\newblock Quantum {HF}/{DFT}-embedding algorithms for electronic structure
  calculations: Scaling up to complex molecular systems.
\newblock {\em The Journal of Chemical Physics}, 154(11):114105, March 2021.

\bibitem{Yalouz_2021}
Saad Yalouz, Bruno Senjean, Jakob G\"{u}nther, Francesco Buda, Thomas~E
  O'Brien, and Lucas Visscher.
\newblock A state-averaged orbital-optimized hybrid
  quantum{\textendash}classical algorithm for a democratic description of
  ground and excited states.
\newblock {\em Quantum Science and Technology}, 6(2):024004, January 2021.

\bibitem{Roos1980}
Bj\"{o}rn~O. Roos, Peter~R. Taylor, and Per~E.M. Sigbahn.
\newblock A complete active space {SCF} method ({CASSCF}) using a density
  matrix formulated super-{CI} approach.
\newblock {\em Chemical Physics}, 48(2):157--173, May 1980.

\bibitem{olsen11}
Jeppe Olsen.
\newblock The {CASSCF} method: A perspective and commentary.
\newblock {\em International Journal of Quantum Chemistry}, 111(13):3267--3272,
  May 2011.

\bibitem{PhysRevB.98.085118}
Nick~S. Blunt, Ali Alavi, and George~H. Booth.
\newblock Nonlinear biases, stochastically sampled effective hamiltonians, and
  spectral functions in quantum monte carlo methods.
\newblock {\em Physical Review B}, 98(8), August 2018.

\bibitem{pyscf1}
Qiming Sun, Timothy~C. Berkelbach, Nick~S. Blunt, George~H. Booth, Sheng Guo,
  Zhendong Li, Junzi Liu, James~D. McClain, Elvira~R. Sayfutyarova, Sandeep
  Sharma, Sebastian Wouters, and Garnet Kin-Lic Chan.
\newblock {PySCF}: the python-based simulations of chemistry framework.
\newblock {\em WIREs Computational Molecular Science}, 8(1):e1340, 2018.

\bibitem{pyscf2}
Qiming Sun, Xing Zhang, Samragni Banerjee, Peng Bao, Marc Barbry, Nick~S.
  Blunt, Nikolay~A. Bogdanov, George~H. Booth, Jia Chen, Zhi-Hao Cui, Janus~J.
  Eriksen, Yang Gao, Sheng Guo, Jan Hermann, Matthew~R. Hermes, Kevin Koh,
  Peter Koval, Susi Lehtola, Zhendong Li, Junzi Liu, Narbe Mardirossian,
  James~D. McClain, Mario Motta, Bastien Mussard, Hung~Q. Pham, Artem Pulkin,
  Wirawan Purwanto, Paul~J. Robinson, Enrico Ronca, Elvira~R. Sayfutyarova,
  Maximilian Scheurer, Henry~F. Schurkus, James E.~T. Smith, Chong Sun,
  Shi-Ning Sun, Shiv Upadhyay, Lucas~K. Wagner, Xiao Wang, Alec White,
  James~Daniel Whitfield, Mark~J. Williamson, Sebastian Wouters, Jun Yang,
  Jason~M. Yu, Tianyu Zhu, Timothy~C. Berkelbach, Sandeep Sharma, Alexander~Yu.
  Sokolov, and Garnet Kin-Lic Chan.
\newblock Recent developments in the {PySCF} program package.
\newblock {\em The Journal of Chemical Physics}, 153(2):024109, 2020.

\bibitem{nielsen_chuang_2010}
Michael~A. Nielsen and Isaac~L. Chuang.
\newblock {\em Quantum Computation and Quantum Information: 10th Anniversary
  Edition}.
\newblock Cambridge University Press, 2010.

\bibitem{on_cnot_cost}
Vivek~V. Shende and Igor~L. Markov.
\newblock On the cnot-cost of toffoli gates.
\newblock {\em Quantum Info. Comput.}, 9(5):461–486, May 2009.

\bibitem{EndoExp}
Suguru Endo, Simon~C. Benjamin, and Ying Li.
\newblock Practical quantum error mitigation for near-future applications.
\newblock {\em Physical Review X}, 8(3), July 2018.

\bibitem{DigitalZNE}
Tudor Giurgica-Tiron, Yousef Hindy, Ryan LaRose, Andrea Mari, and William~J.
  Zeng.
\newblock Digital zero noise extrapolation for quantum error mitigation.
\newblock 2020 IEEE International Conference on Quantum Computing and
  Engineering (QCE), Denver, CO, USA, 2020, 2020.

\bibitem{Cincio2018}
Lukasz Cincio, Yiǧit Subaşi, Andrew~T. Sornborger, and Patrick~J. Coles.
\newblock {Learning the quantum algorithm for state overlap}.
\newblock {\em New Journal of Physics}, 20(11), 2018.

\bibitem{HEAAnsatz}
Abhinav Kandala, Antonio Mezzacapo, Kristan Temme, Maika Takita, Markus Brink,
  Jerry~M. Chow, and Jay~M. Gambetta.
\newblock Hardware-efficient variational quantum eigensolver for small
  molecules and quantum magnets.
\newblock {\em Nature}, 549(7671):242--246, September 2017.

\bibitem{Rotosolve}
Mateusz Ostaszewski, Edward Grant, and Marcello Benedetti.
\newblock Structure optimization for parameterized quantum circuits.
\newblock Quantum 5, 391 (2021), 2019.

\bibitem{PhysRevB.102.075104}
Alexey Uvarov, Jacob~D. Biamonte, and Dmitry Yudin.
\newblock Variational quantum eigensolver for frustrated quantum systems.
\newblock {\em Physical Review B}, 102(7), August 2020.

\bibitem{wang2021noiseinduced}
Samson {Wang}, Enrico {Fontana}, M.~{Cerezo}, Kunal {Sharma}, Akira {Sone},
  Lukasz {Cincio}, and Patrick~J. {Coles}.
\newblock {Noise-Induced Barren Plateaus in Variational Quantum Algorithms},
  July 2020.

\bibitem{Grant2019initialization}
Edward Grant, Leonard Wossnig, Mateusz Ostaszewski, and Marcello Benedetti.
\newblock An initialization strategy for addressing barren plateaus in
  parametrized quantum circuits.
\newblock {\em Quantum}, 3:214, December 2019.

\bibitem{Adaptvqe}
Harper~R. Grimsley, Sophia~E. Economou, Edwin Barnes, and Nicholas~J. Mayhall.
\newblock An adaptive variational algorithm for exact molecular simulations on
  a quantum computer.
\newblock {\em Nature Communications}, 10(1), July 2019.

\end{thebibliography}


\end{document}